\documentclass[aps,pre,groupedaddress,10pt,notitlepage]{revtex4-1}

\usepackage{amsmath}
\usepackage{amssymb}
\usepackage{amsthm}
\usepackage{bbm}
\usepackage{graphicx}
\usepackage{float}
\usepackage{afterpage}
\usepackage{placeins}
\usepackage{epstopdf}

\usepackage{hyperref}
\usepackage[usenames,dvipsnames]{xcolor}
\hypersetup{colorlinks=true, linkcolor=BrickRed, urlcolor=blue!50!black, citecolor=blue!50!black}

\newcommand\diff{\mathrm{d}}
\renewcommand{\vec}[1]{\mathbf{#1}}
\renewcommand{\imath}[0]{\mathsf{i}}

\begin{document}
\title{Intermediate scattering function of an anisotropic active Brownian particle}
\author{Christina Kurzthaler}
\affiliation{Institut f\"ur Theoretische Physik, Universit\"at Innsbruck, Technikerstra{\ss}e 21A, A-6020 Innsbruck, Austria}
\author{Sebastian Leitmann}
\affiliation{Institut f\"ur Theoretische Physik, Universit\"at Innsbruck, Technikerstra{\ss}e 21A, A-6020 Innsbruck, Austria}
\author{Thomas Franosch}
\email[]{thomas.franosch@uibk.ac.at}
\affiliation{Institut f\"ur Theoretische Physik, Universit\"at Innsbruck, Technikerstra{\ss}e 21A, A-6020 Innsbruck, Austria}
\date{\today}

\begin{abstract}
Various challenges are faced when animalcules such as bacteria, protozoa,
  algae, or sperms move autonomously in aqueous media at low Reynolds number.
  These active agents are subject to strong stochastic fluctuations, that
  compete with the directed motion. So far most studies consider the lowest
  order moments of the displacements only, while more general spatio-temporal information on
  the stochastic motion is provided in scattering experiments. Here we derive
  analytically exact expressions for the directly measurable intermediate
  scattering function for a mesoscopic model of a single, anisotropic 
  active Brownian particle in three dimensions. The mean-square displacement and the non-Gaussian
  parameter of the stochastic process are obtained as derivatives of the
  intermediate scattering function. These display different temporal
  regimes dominated by effective diffusion and directed motion due to the
  interplay of translational and rotational diffusion which is rationalized
  within the theory. The most prominent feature of the intermediate scattering
  function is an oscillatory behavior at intermediate wavenumbers reflecting
  the persistent swimming motion, whereas at small length scales bare
  translational and at large length scales an enhanced effective diffusion
  emerges.  We anticipate that our characterization of the motion of
  active agents will serve as a reference for more realistic models and experimental observations. 
\end{abstract}

\maketitle

\section*{Introduction}
Active particles are intrinsically out of equilibrium and exhibit
peculiar dynamical behavior ~\cite{Romanczuk:2012,Vicsek:2012, Marchetti:2013,
Elgeti:2015, Bechinger:2016} on the single as well as on the collective level.
These active agents are ubiquitous in nature and include bacteria
~\cite{Berg:1972,Berg:1990,Lauga:2006,Copeland:2009},
algae~\cite{Merchant:2007}, unicellular
protozoa~\cite{Machemer:1972,Blake:1974,Roberts:2010}
or spermatozoa~\cite{Woolley:2003,Riedel:2005}, that move due to a single or an
array of flagella pushed by molecular motors. Only recently, artificial active
particles have been synthesized and are self-propelled by either biomimetic
motors ~\cite{Dreyfus:2005,Kudrolli:2010}, or due to the response of their patterned surface
to chemical or temperature gradients, thereby converting chemical energy into
directed motion
~\cite{Howse:2007,Jiang:2010,Zheng:2013,tenHagen:2014,Lee:2014}.  Furthermore,
they also move in crowded media and their effective swimming speed is
strongly determined by the viscoelasticity and geometrical constraints of the
surroundings~\cite{Martinez:2014,Brown:2016}.  

To capture analytically the intricacies of the propulsion mechanisms, simple
models for single swimmers have been conceived on different levels of
coarse-graining. Microscopic theories for squirmers
\cite{Lighthill:1952,Blake:1971}, linked-bead swimmers
\cite{Golestanian:2004,Felderhof:2015, Smith:2015},  self-thermophoresis
\cite{Jiang:2010}, and, self-diffusiophoresis~\cite{Wuerger:2015} of Janus
particles  have been elaborated and include the full hydrodynamic flow.  On a
larger scale, effective models for individual self-propelled particles ignoring
hydrodynamics and the origin of the swimming motion are used to describe the
stochastic motion and the dynamic behavior.   There, the dynamics is modeled in
terms of non-equilibrium Langevin equations~\cite{Romanczuk:2012,tenHagen:2014,
Sevilla:2014, vanTeffelen:2008} such that the noise strength is an effective
parameter unrelated to the temperature of the environment, in striking contrast
to the fluctuation-dissipation theorem for equilibrium dynamics.  In
particular, these equations of motion serve as a suitable starting point for
simulations~\cite{Volpe:2014}. 

The complexity of these transport properties has often been quantified
experimentally and in simulations in terms of low-order moments of the
displacements ~\cite{Howse:2007,Zheng:2013,Brown:2016} and compared to
theoretical models.  For example, generically the mean-square displacement
exhibits a regime resembling ballistic motion which directly reflects the
persistent swimming.  Only at longer times the motion becomes randomized and
the mean-square displacement increases as anticipated from conventional
diffusion.  Higher moments can be derived \cite{Zheng:2013} in principle from
the stochastic equtions of motion, yet the calculations become more and more
cumbersome with increasing order. However, these low-order moments provide
only restricted information on the statistical properties of the random
displacements as a function of time, in particular, they are to a large extend
insensitive to the shape of the probability distribution.  

More general spatio-temporal information is encoded in the intermediate
scattering function $F(k,t)$, which resolves the motion of the particle at
lag-time $t$ on a length scale $2\pi/k$, and is directly measurable in
scattering experiments~\cite{Berne:1976} such as dynamic light scattering. The
same quantity can be obtained by advanced image analysis within the recently
developed differential dynamic microscopy (DDM)~\cite{Martinez:2012,Poon:2016},
which provides direct access to the relevant length scales of active particles.
Of course, single-particle tracking also collects the full statistical
information and the intermediate scattering function can be obtained from this information, 
yet often the temporal resolution is not high enough to monitor the dynamics on
small length scales. Last, the intermediate scattering function can also be
viewed as the characteristic function~\cite{Gardiner:2009} of the random displacements, which is
equivalent to the full probability distribution. In particular, the moments of
the displacements are encoded as derivatives with respect to the wavenumber.
Theoretical approaches to the intermediate scattering function for active
particles are rare \cite{Sevilla:2015} and no exact solutions appear to be
available.  

\section*{Dynamics of an Active Brownian Particle \label{sec:langevin}}  
\subsection*{Model}

\begin{figure}[h]
\centering
\includegraphics[width=0.5\linewidth, keepaspectratio]{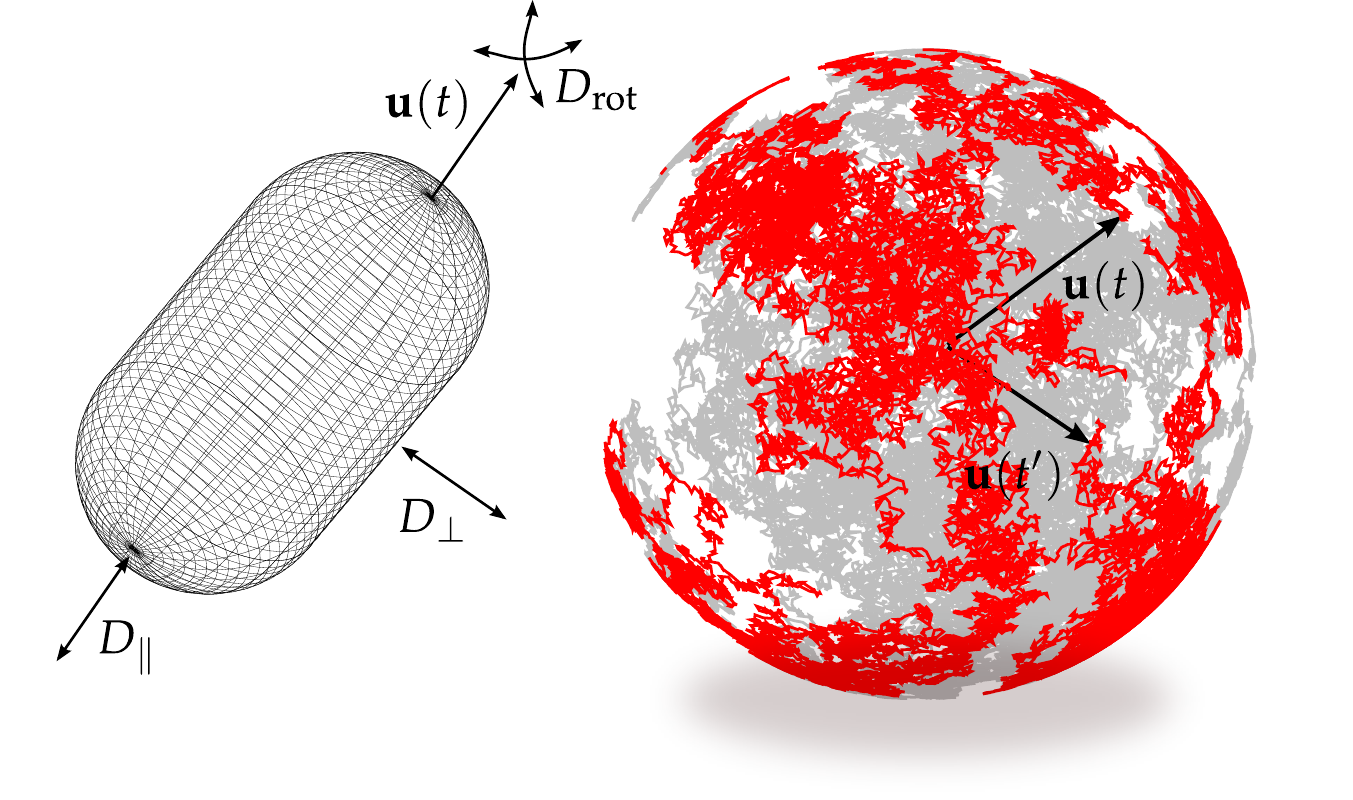}
\caption{Model set up. Left: Anisotropic particle with orientation
$\vec{u}(t)$ and translational $D_\parallel,D_\perp$ and rotational
$D_\text{rot}$ diffusion coefficients.  Right: Diffusion of the orientation
$\vec{u}(t)$ on the unit sphere. \label{fig:rod}} 
\end{figure}

We assume the active Brownian particle to move at constant velocity $v$ along its
instantaneous orientation $\vec{u}(t)$ subject to random fluctuations
determined by the rotational diffusion coefficient $D_\text{rot}$.  This
diffusion process can geometrically be regarded as the diffusion of the
orientation $\vec{u}(t)$ on the unit sphere, as Fig.~\ref{fig:rod}. 
In addition, the motion of the anisotropic active particle is
characterized by axisymmetric translational diffusion measured in terms of
the short time diffusion coefficients parallel ($D_\parallel$) and
perpendicular ($D_\perp$) to the anisotropic particle, Fig.~\ref{fig:rod}.  
\noindent 
Hence, for a three dimensional swimmer the dynamics are described by the
Langevin equations in It$\bar{\text{o}}$ form for the position $\vec{r}(t)$ and
the orientation $\vec{u}(t)$
\begin{align}
  \diff \vec{u}(t) &= -2D_\text{rot}\vec{u}(t) \diff t-\sqrt{2D_\text{rot}}\vec{u}(t)\times \diff\boldsymbol{\xi}(t), \label{eq:u}\\
  \diff \vec{r}(t) &= v\vec{u}(t) \diff t + \left[\sqrt{2D_\parallel}\vec{u}(t)\vec{u}(t)^{\text{T}}+\sqrt{2D_\perp}\left(\mathbb{I}-\vec{u}(t)\vec{u}(t)^\text{T}\right)\right]\diff\boldsymbol{\zeta}(t).\label{eq:r}
\end{align}
Here the diffusion coefficients $D_\parallel$ and $D_\perp$ for the motion
along and perpendicular to the axis of the swimmer encode the
translational-rotational coupling.  The random fluctuations are modeled in
terms of independent white noise processes, $\boldsymbol \xi(t)$ and
$\boldsymbol \zeta(t)$ with zero mean and covariance $\langle
\xi_i(t)\xi_j(t')\rangle =
\langle\zeta_i(t)\zeta_j(t')\rangle=\delta_{ij}\delta(t-t')$ for $i,j=1,2,3$.
The drift term in Eq.~(\ref{eq:u}) ensures that the normalization
condition remains fulfilled, $\diff[\vec{u}(t)^2]/\diff t =0$.  Let us
emphasize that if the Stratonovich interpretation is used, the drift term
in the equation for the orientation needs to be dropped. 

The model contains two dimensionless parameters, first the translational
anisotropy $\Delta D = D_\parallel-D_\perp$ relative to the mean diffusion
coefficient $\bar{D}=(D_\parallel+2D_\perp)/3$.  For passive rod-like particles
in the limit of very large aspect ratio
hydrodynamic suggests $D_\parallel = 2D_\perp$~\cite{Doi:1986}, such that $\Delta
D/\bar{D}=3/4$. Here we consider $D_\parallel$ and $D_\perp$ as effective
parameters quantifying the noise only, and the anisotropy can take arbitrary
values in $-3/2 \leq \Delta D/\bar{D} \leq 3$. 
Next, the problem displays a characterstic length,
$a=\sqrt{3\bar{D}/D_\text{rot}}/2$, which corresponds to the geometric radius of a spherical
particle in the case of equilibrium diffusion coefficients $D_\text{rot}=k_\text{B}T/8\pi\eta a^3$ 
and $\bar{D}=k_\text{B}T/6\pi\eta a$. Then the second dimensionless
parameter is the P{\'e}clet number $\text{Pe}=va/\bar{D}$ 
measuring the relative importance of the active motion with respect to
diffusion.
\subsection*{Analytic solution}
From the stochastic differential equations one derives the Fokker-Planck
equation~\cite{Birkhaeuser:2009,Gardiner:2009} for the time
evolution of the probability density
$\mathbb{P}(\vec{r},\vec{u},t|\vec{r}_0,\vec{u}_0,t_0)$ to find the swimmer at
position $\vec{r}$, with orientation $\vec{u}$ at time $t$ given that it has
been at some position $\vec{r}_0$ with initial orientation $\vec{u}_0$ at an
earlier time $t_0$. Since the stochastic process is translational invariant in
time and space, only displacements $\Delta \vec{r}=\vec{r}-\vec{r}_0$ and
lag times $t$ (with $t_0=0$) have to be considered,
$\mathbb{P}\equiv\mathbb{P}(\Delta\vec{r},\vec{u},t|\vec{u}_0)$.  Then the
Fokker-Planck equation assumes the form
\begin{align}
  \partial_t \mathbb{P} &= -v\vec{u}\cdot\partial_\vec{r}\mathbb{P}+D_\text{rot} \Delta_{\vec{u}}\mathbb{P}+\partial_\vec{r}\cdot(\vec{D}\cdot\partial_\vec{r}\mathbb{P}),\label{eq:FP}
\end{align} 
subject to the initial condition $\mathbb{P}(\Delta
\vec{r},\vec{u},t=0|\vec{u}_0)=\delta(\Delta\vec{r})\delta^{(2)}(\vec{u},\vec{u}_0)$, 
where the delta function on the surface of the sphere $\delta^{(2)}(\cdot,\cdot)$
enforces both orientations to coincide.
Here, $\partial_\vec{r}$ denotes the spatial gradient, $\Delta_\vec{u}$ the
angular part of the Laplacian, reflecting the orientational diffusion, and
$\vec{D}=D_\parallel\vec{u}\vec{u}^\text{T}+D_\perp(\mathbb{I}-\vec{u}\vec{u}^\text{T})$.
The first term on the right describes the active motion, in addition to the
standard Smoluchowski-Perrin equation~\cite{Doi:1986} for the diffusion of an
anisotropic particle.  The Fokker-Planck equation for $\mathbb{P}$ simplifies
upon a spatial Fourier transform 
\begin{align}
  \widetilde{\mathbb{P}}(\vec{k},\vec{u},t|\vec{u}_0)& = \int\!\diff^3 r \exp(-\imath\vec{k}\cdot\vec{r})\mathbb{P}(\vec{r},\vec{u}, t|\vec{u}_0),\label{eq:Fourier} 
\end{align} 
which solves the equation of motion
\begin{align}
  \partial_t \widetilde{\mathbb{P}} &= D_\text{rot}\Delta_\vec{u}\widetilde{\mathbb{P}} -\imath v \vec{u}\cdot\vec{k}\widetilde{\mathbb{P}} -[D_\perp \vec{k}^2+\Delta D(\vec{u}\cdot\vec{k})^2]\widetilde{\mathbb{P}}.\label{eq:PDEchar}
\end{align}
The quantity of interest in scattering experiments~\cite{Berne:1976} is the
intermediate scattering function (ISF) 
\begin{align}
  F(\vec{k},t)	&= \langle \exp[-\imath \vec{k}\cdot\Delta\vec{r}(t)]\rangle,\label{eq:ISF}  
\end{align} 
which is obtained by marginalizing over all final orientations $\vec{u}$ and
averaging over all initial orientations $\vec{u}_0$,
\begin{align}
  F(\vec{k},t)&= \int\!\diff^2 u\!\int\!\frac{\diff^2u_0}{4\pi} \ \widetilde{\mathbb{P}}(\vec{k},\vec{u},t|\vec{u}_0). 
\end{align}
The ISF can also be interpreted as the characteristic
function~\cite{Gardiner:2009} of the random displacement variable $\Delta
\vec{r}(t)$.   In particular, the moments are obtained by taking derivatives
with respect to the wave vector $\vec{k}$. Since after averaging the motion is
isotropic, the ISF $ F(k,t) \equiv F(\vec{k},t)$ depends only on the magnitude of the wave vector
$k=|\vec{k}|$. Averaging over the directions of $\vec{k}$ yields the equivalent representation 
\begin{align}
  F(k,t) &=\left\langle \frac{\sin(k|\Delta\vec{r}(t)|)}{k|\Delta\vec{r}(t)|}\right\rangle \label{eq:sinc}
\end{align}
and the expansion of the ISF for small wavenumbers 
\begin{align}
  F(k,t) &= 1- \frac{k^2}{3!}\langle|\Delta\vec{r}(t)|^2\rangle+\frac{k^4}{5!}\langle|\Delta\vec{r}(t)|^4\rangle+\mathcal{O}(k^6)\label{eq:expansionSinc}
\end{align}
allows one to recover the mean-square displacement
$\langle|\Delta\vec{r}(t)|^2\rangle$ and the mean-quartic displacement $\langle
|\Delta\vec{r}(t)|^4\rangle$ by comparing the corresponding terms in the small wavenumber
expansion. More generally, even moments can be obtained numerically by taking derivatives 
of the ISF with respect to the squared wavenumber, 
\begin{align}
\langle |\Delta \vec{r}(t)|^{2n}\rangle &= (-1)^n \frac{(2n+1)!}{n!} \left. \frac{\partial^n}{\partial (k^2)^n} F(k,t)\right|_{k^2=0}.
\end{align}
The equation of motion Eq.~(\ref{eq:PDEchar})  is reminiscent 
of a Schr\"odinger equation on the unit sphere and can be solved 
by separation of variables. We parametrize the orientation 
 $\vec{u}=(\sin\vartheta \cos\varphi,\sin\vartheta\sin\varphi,\cos\vartheta)^T$ in terms
of its polar angles, and similarly for $\vec{u}_0$.
Then the solution is a superposition 
of appropriate eigenfunctions
\begin{align}
  \widetilde{\mathbb{P}}(\vec{k},\vec{u},t|\vec{u}_0)=\frac{1}{2\pi}e^{-D_\perp k^2 t}\sum_{\ell=0}^\infty\sum_{m=-\infty}^\infty\! e^{\imath m (\varphi-\varphi_0)}\text{Ps}_\ell^m(c,R,\eta)\text{Ps}_\ell^m(c,R,\eta_0)e^{-A^m_{\ell} D_\text{rot}t}.\label{eq:expansionP}
\end{align} 
Here we abbreviated $\eta = \cos\vartheta$, $\eta_0 = \cos\vartheta_0$, and $\text{Ps}_\ell^m(c,R,\eta)$ are 
the generalized spheroidal wave functions of order $m$ and degree $\ell$ \cite{Yan:2009,NIST:online, NIST:print}.
They solve the corresponding eigenvalue problem 
\begin{align}
  \left[\frac{\diff}{\diff\eta}\left((1-\eta^2)\frac{\diff}{\diff \eta}\right)+R\eta-c^2\eta^2-\frac{m^2}{1-\eta^2}+A^m_{\ell}\right]\text{Ps}_\ell^m(c,R,\eta)&= 0,\label{eq:Pslm}
\end{align}
with eigenvalue $A^m_{\ell}=A^m_{\ell}(R,c)$ and we identify the dimensionless parameters $R = - \imath kv/D_\text{rot}$ 
and $c^2= \Delta D k^2/D_\text{rot}$. Hence,
at fixed wavenumber $k$, $R$ parametrizes the importance of active motion with
respect to orientational diffusion whereas $c$ measures the coupling of the
translational and orientational diffusion. In particular the ratio
$|R/c|=\text{Pe}\sqrt{4\Delta D/3\bar{D}}$ is wavenumber-independent.
 
Integrating Eq.~(\ref{eq:expansionP}) over the polar angles, only $\text{Ps}_\ell^0$ contributes and we obtain  
\begin{align}
F(k,t) &=\frac{1}{2}e^{-D_\perp k^2 t}\sum_{\ell=0}^\infty e^{-D_\text{rot}A^0_{\ell}t}\Bigl[\int_{-1}^1 \diff\eta \text{Ps}_\ell^0(c,R,\eta) \Bigr]^2,\label{eq:solISF}
\end{align}
The explicit expression  Eq.~(\ref{eq:solISF}) for the intermediate scattering
function $F(k,t)$ in terms of the generalized spheroidal wave functions is one
of the principal results of this work.  

\subsection*{Exact low moments} 
The low-order moments can be obtained upon expanding the ISF for small wave
numbers (Eq.~(\ref{eq:solISF})) such that the moments can be identified with Eq.~(\ref{eq:expansionSinc}).
Here we illustrate the derivation only for the mean-square displacement. 

For $R=0$ and $c^2=0$ the spheroidal wave functions reduce to the Legendre
polynomials, $\text{Ps}_\ell^0(0,0,\eta) =
\text{P}_\ell(\eta)\sqrt{(2\ell+1)/2}$ with eigenvalues
$A^0_{\ell}(0,0)=\ell(\ell+1)$.  For small dimensionless parameters $R$, $c$ the
Legendre polynomials are deformed analytically, to order $\mathcal{O}(k^2)$,  as 
required for the mean-square displacement Eq.~(\ref{eq:expansionSinc}), 
the $\text{Ps}_\ell^0$ acquire contributions $\text{P}_\ell$,
$\text{P}_{\ell\pm1}$, and, $\text{P}_{\ell\pm2}$, concomitantly the eigenvalues $A^0_{\ell}$ shift.  
The explicit expressions are lengthy and deferred to the methods section.  
The integral in Eq.~(\ref{eq:solISF}) can then be performed using the orthogonality
of the Legendre polynomials and one concludes that only terms $\ell \leq 2$ need
to be taken into account to order $\mathcal{O}(k^2)$. Yet, inspection of Eq.~(\ref{eq:Psexpansion})
of the methods section shows that integration of $\text{Ps}_2^0(R,c,\eta)$ yields terms of 
order $\mathcal{O}(R^2)$ and $\mathcal{O}(c^2)$ and after squaring in Eq.~(\ref{eq:solISF}) 
of only order $\mathcal{O}(k^4)$. 
Hence, the contributing eigenfunctions for the mean-square displacement evaluate to 
\begin{align}
  \frac{1}{\sqrt{2}}\int_{-1}^1\diff\eta\text{Ps}_\ell^0(R,c,\eta) &=  
  \begin{cases}
    1-R^2/24   +\mathcal{O}(\cdot)   & \text{ } \ell=0, \\
    -R/2\sqrt{3}+\mathcal{O}(\cdot)  & \text{ } \ell=1.\\
  \end{cases}
\end{align}
and the corresponding eigenvalues read
\begin{align}
  A^0_{\ell}(R,c) &=  \begin{cases}
  c^2/3 - R^2/6 +\mathcal{O}(\cdot),     & \text{ } \ell=0,\\
  2+3c^2/5 +R^2/10+\mathcal{O}(\cdot),  & \text{ } \ell=1.\\
                       \end{cases}
\end{align}
Collecting results for the ISF $F(k,t)$ to order $\mathcal{O}(k^2)$ and comparing with Eq.~(\ref{eq:expansionSinc}),
yields for the mean-square displacement
\begin{align}
  \langle|\Delta\vec{r}(t)|^2\rangle &=  \frac{v^2}{2D_\text{rot}^2}(e^{-2D_\text{rot}t}+2D_\text{rot}t-1) +6\bar{D}t.\label{eq:msd}
\end{align}
This expression generalizes the earlier result for the case of an isotropic
active agent~\cite{Sevilla:2014,Sevilla:2015} and anisotropic passive
particle~\cite{Doi:1986,Han:bla:2006}. It also recovers the mean-square displacement of a
freely rotating ellipsoidal particle~\cite{tenHagen:2011} obtained directly
from the Langevin equations. Alternatively $\langle |\Delta\vec{r}(t)|^2\rangle$
can be calculated by time-dependent perturbation theory from Eq.~(\ref{eq:PDEchar}) up to second
order.  

The first contribution to the mean-square displacement in Eq.~(\ref{eq:msd}) reflects the
active motion, which displays directed motion $v^2t^2$ for times $t \lesssim
\tau_\text{rot}:=D_\text{rot}^{-1}$ where the particle does not
change its direction significantely. During this time the particle
covers a typical distance $L=v/D_\text{rot}$, which we refer to as the
persistence length.  In contrast at times $t \gtrsim \tau_\text{rot}$ the
active contribution increases linearly $v^2 t/6D_\text{rot}$ where the
orientational degree of freedom is relaxed. The second contribution is merely
the isotropically averaged translational motion. Interestingly at the level of
the mean-square displacement there is no coupling between the translational
diffusion and the active motion induced by the orientational diffusion. 
\begin{figure}[htp]
\centering
\includegraphics[width = \linewidth, keepaspectratio]{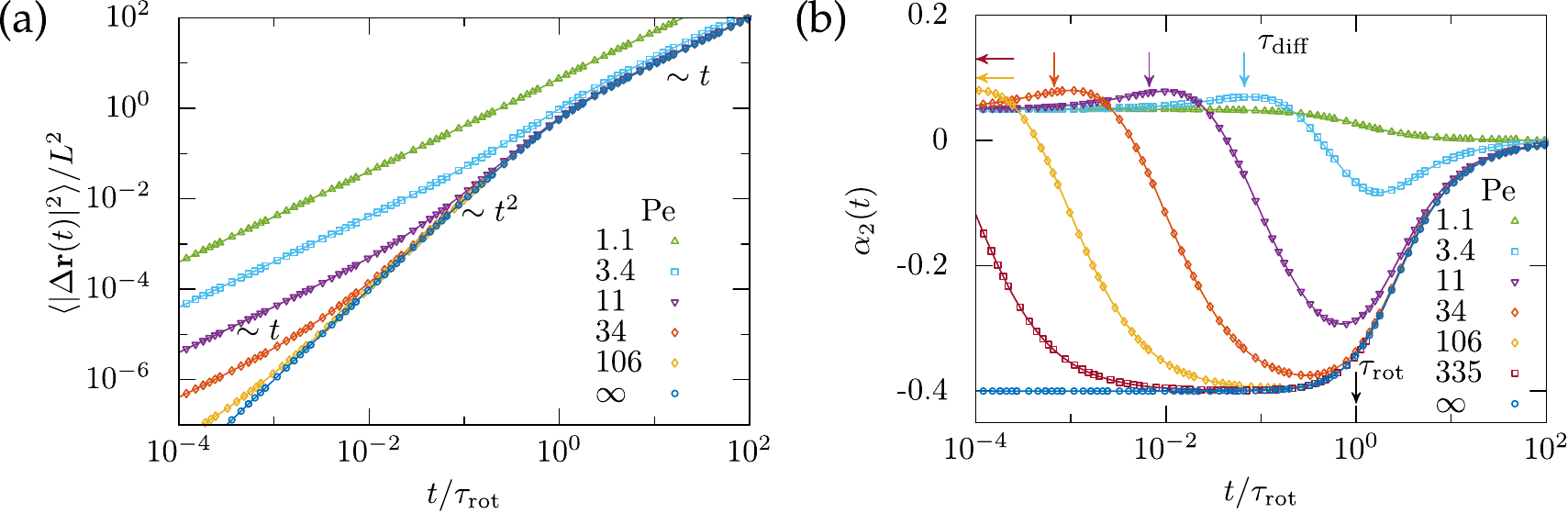}
\caption{Exact low-order moments of a single
   self-propelled particle subject to translational Brownian motion with hydrodynamic anisotropy $\Delta
     D/\bar{D} = 3/4$. (a) Mean-square displacement $\langle |\Delta\vec{r}(t)|^2\rangle/L^2$ in
   units of the persistence length $L=v/D_\text{rot}$, and, (b) non-Gaussian parameter $\alpha_2(t)$ for 
   different P{\'e}clet numbers, $\text{Pe} =va/\bar{D}$.
 Simulation and theory results are shown using symbols and lines,
 respectively.
   \label{fig:moments}}
\end{figure}\noindent

From the mean-square displacement we identify three temporal windows,
Fig.~\ref{fig:moments}~(a).  For short times $t \lesssim \tau_\text{diff}:=\bar{D}/v^2$
it increases linearly by the translational diffusion only, while at longer
times the persistent swimming motion dominates.  At even longer times $t
\gtrsim \tau_\text{rot}$ the mean-square displacement increases again linearly
with an effective diffusion coefficient $D_\text{eff} =
\bar{D}+v^2/6D_\text{rot}$, equivalently the enhancement is
$D_\text{eff}/\bar{D}=1+2\text{Pe}^2/9$.   
The crossover from persistent motion to
effective diffusion occurs at length scale
$L^2[1+\mathcal{O}(\text{Pe}^{-2})]$.  The window of persistent motion is set
by the ratio of the two crossover times
$\tau_\text{rot}/\tau_\text{diff}=4\text{Pe}^2/3$ and opens upon increasing the
P{\'e}clet number. 

Extending the expansion  of the intermediate scattering function up to fourth
order in the wavenumber $k$ is tedious and the result is lengthy, 
\begin{align}
  \langle|\Delta\vec{r}(t)|^4\rangle &= \Bigl[\bigl\{8 D_\text{rot}^2 [405 D_\text{rot}^2 \bar{D}^2t^2+2 \Delta D^2(6 D_\text{rot} t-1)]
                        +4D_\text{rot}v^2[135D_\text{rot}\bar{D}t(2D_\text{rot}t-1)+\Delta D (60D_\text{rot}t-52)]\notag\\
& \ \ \ \ \ +v^4[107+6D_\text{rot}t(15D_\text{rot}t-26)]\bigr\}+18e^{-2D_\text{rot}t}v^2\bigl( 30D_\text{rot}^2\bar{D}t+4\Delta D D_\text{rot}(3+2D_\text{rot}t)-3v^2(2+D_\text{rot}t)\bigr)\notag\\
& \ \ \ \ \ +e^{-6D_\text{rot}t}(v^2-4\Delta D D_\text{rot})^2\Bigr]/54D_\text{rot}^4.
\end{align}
In contrast to the mean-square displacement, the mean-quartic displacement
depends explicitly on the translational anisotropy $\Delta D$ such that the
rotational-translational coupling becomes important. We shall see below that
depending on $\Delta D$ the dynamics becomes qualitatively different.  

Rather then the mean-quartic displacement, we focus on the non-Gaussian
parameter~\cite{Hofling:2013} 
\begin{align} \alpha_2(t)	&=
\frac{3\langle|\Delta\vec{r}(t)|^4\rangle}{5\langle|\Delta\vec{r}(t)|^2\rangle^2}-1,
\end{align} 
which is a sensitive indicator on how far the process deviates from
diffusion, see Fig. \ref{fig:moments}~(b).  

For long times $t\gtrsim \tau_\text{rot}$
the non-Gaussian parameter approaches zero $\mathcal{O}(t^{-1})$ for all P{\'e}clet numbers as anticipated
by the central limit theorem. Interestingly, for the limiting case of a self-propelled
particle without any translational diffusion, $\text{Pe} = \infty$, one infers
$\alpha_2(t\to 0)= -2/5$, which reflects the persistent swimming motion at
short-times. In contrast, for non-vanishing translational diffusion, 
$\text{Pe}<\infty$,  
the non-Gaussian parameter approaches a constant
$\alpha_2(t\lesssim\tau_\text{diff})= 4\Delta D^2/45\bar{D}^2$ for
short-times, as anticipated
for anisotropic translational diffusion.  
In particular, for $D_\parallel=2D_\perp$ it assumes the value $\alpha_2(t\lesssim\tau_\text{diff})=1/20$, 
whereas it vanishes for isotropic diffusion.
For large P{\'e}clet number
there is an extended intermediate temporal regime, 
where the non-Gaussian parameter is close to the one for infinite P{\'e}clet number, 
thereby, a prominent minimum emerges. Here the negative non-Gaussian
parameter can be traced back to the directed swimming motion, which 
dominates the translational diffusion of the active agent at these intermediate times.
Thus, for decreasing $\tau_\text{diff}$ the intermediate negative
plateau of directed swimming motion in the non-Gaussian parameter is observed
for longer times, see Fig.~\ref{fig:moments}~(b).

For the parameters shown in Fig. \ref{fig:moments}~(b) an additional
maximum occurs at shorter times.  One can work out analytically from the initial 
slope of $\alpha_2(t)$ that this
happens only for positive anisotropies $\Delta D > 0$ and P{\'e}clet numbers
$\text{Pe}> \sqrt{3 \Delta D/2D_\perp}$.  Conversely, we conclude that a maximum in the
non-Gaussian parameter is a genuine fingerprint of active motion.  

\subsection*{Intermediate scattering function} 
\begin{figure*}[htp]
 \centering
 \includegraphics[width=\linewidth, keepaspectratio]{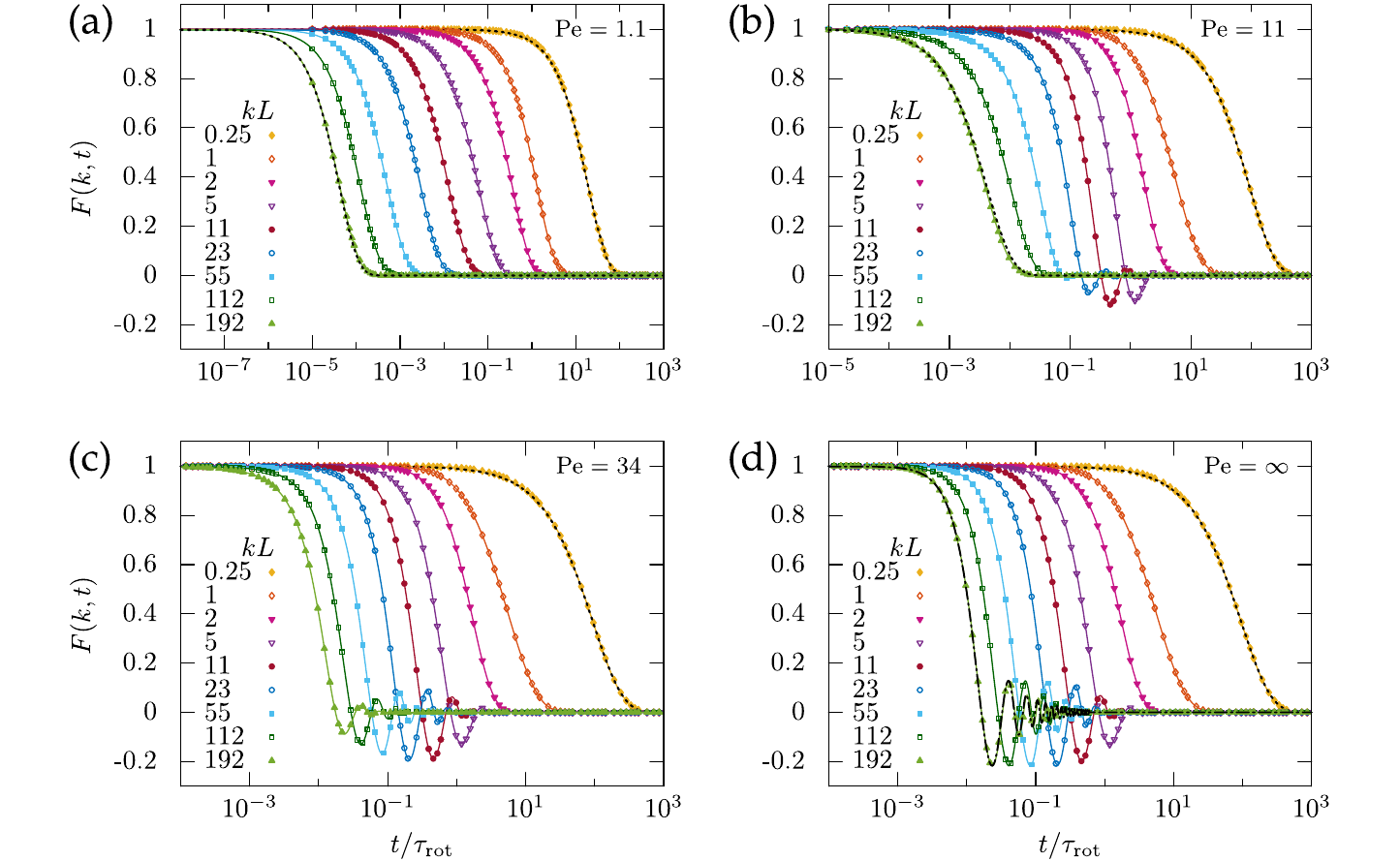}\\
 \caption{Intermediate scattering function $F(k,t)$ of an active
    Brownian particle subject to translational diffusion (here $\Delta D /\bar{D}=3/4$) 
  for the full range of wavenumbers $k$ measured in terms of the persistence length $L=v/D_\text{rot}$.  
  The dashed line represents relaxing exponentials  $\exp(-D_\text{eff}k^2t)$ and $\exp(-\bar{D}k^2t)$ for small
  and large wavenumbers, respectively. The dashed-dotted line indicates the sinc function $\sin(kvt)/kvt$. 
 \label{fig:corr}}
 \end{figure*}

We have evaluated numerically the series for the intermediate scattering
function in Eq.~(\ref{eq:solISF}) for arbitrary times and wavenumbers and compare the results
to stochastic simulations, see Fig.~\ref{fig:corr}.  The natural scale for the wavenumbers
$k$ is set by the persistence length $L$, and our data cover the small length
scales resolving the persistent swimming motion as well as large length scales
where the particle undergoes a random walk.  Indeed for small wavenumbers the
ISF are well approximated by an effective diffusion, $\exp(-D_\text{eff}k^2t)$
with the effective diffusion coefficient obtained from the long-time behavior of the
mean-square displacement.  Increasing the wavenumber the qualitative behavior
depends on the P{\'e}clet number. 

For small P{\'e}clet number (see Fig.~\ref{fig:corr}~(a)) the ISF decreases monotonically for all wave
numbers, in particular, the large wavenumbers approach again an exponential
$\exp(-\bar{D}k^2t)$ characterized by the mean translational diffusion
coefficient $\bar{D}$. This behavior is consistent with the linear increase of
the mean-square displacement, Fig.~\ref{fig:moments}~(a), for small P{\'e}clet numbers. For
intermediate wavenumbers (Fig.~\ref{fig:corr}~(b)-(c)) the shape of the ISF is no longer a pure exponential
since the rotational-translational coupling becomes relevant at time scales
$t\lesssim\tau_\text{rot}$. 

For P{\'e}clet numbers, $\text{Pe}\gtrsim 3.4$, the ISF displays
damped oscillations for wavenumbers that start to resolve the motion on the scale 
of the persistence length. At length scales $\bar{D}k^2\tau_\text{diff}\gtrsim 1$
short time diffusion takes over again, see Fig.~\ref{fig:corr}~(b). Inserting the definition of 
$\tau_\text{diff}$, one infers that this regime corresponds to length scales $ka \gtrsim\text{Pe}$
where the swimmer moves only a fraction of its size $a$.
In particular, for high P{\'e}clet numbers $\text{Pe}\gtrsim 12$ the short time diffusion 
is no longer resolved for the wavenumbers shown in Fig.~\ref{fig:corr}~(c).      
For infinite P{\'e}clet number, the translational diffusion is negligible and the ISF
oscillates for wavenumbers resolving the persistence length, Fig.~\ref{fig:corr}~(d). 

The physics of these oscillations can be rationalized easily by inspecting the general
expression of the ISF, Eq.~(\ref{eq:sinc}). For wavenumbers such that the rotational and translational diffusion can be ignored,
the trajectories can be approximated by purely persistent motion $|\Delta \vec{r}(t)|=vt$ and
there the ISF follows $F(k,t)	=\sin(vkt)/vkt,$  
as has been discussed already in Ref. \cite{Berne:1976}. 
For infinite P{\'e}clet number the sinc function serves as a good
approximation for wavenumbers $kL\gtrsim 20$.   

It is also interesting to ask how the oscillations emerge mathematically from the general
solution in terms of eigenfunctions, Eq.~(\ref{eq:solISF}). Naively, one expects that the ISF is a sum of 
relaxing exponentials only, in particular, they should decay monotonically. Yet, the 
operator in Eq.~(\ref{eq:Pslm}) for the eigenvalue problem is non-Hermitian, since $R=-\imath kL$ is not real,  
such that the eigenvalues can become complex. Indeed one can show (see section Methods), for example $\text{Pe}=\infty$, that
at $|R|=1.9$ the two lowest real eigenvalues merge and bifurcate to a pair of complex conjugates.
Further bifurcations for larger eigenvalues occur at even larger $|R|$. For large P{\'e}clet numbers 
the scenario is qualitatively similar, whereas for small $\text{Pe}$ the eigenvalues remain real and no 
oscillations in the ISF emerge. Since the eigenvalues depend non-analytically on $|R|=kL$, there
is a finite radius of convergence for the expansion of the ISF in powers of $k$ set by the 
first bifurcation point. In particular, the oscillations cannot be obtained by extending the series expansion,
 Eq.~(\ref{eq:expansionSinc}), in terms of the moments to arbitrary order.



%
%

%

\section*{Summary and Conclusion}
We have determined exact analytic expressions for the intermediate 
scattering function (ISF) of an anisotropic active Brownian particle 
in terms of an expansion of eigenfunctions. The solution is validated and 
exemplified by stochastic simulations.
Interestingly, the ISF displays a regime 
with oscillatory behavior in striking contrast to passive motion in equilibrium 
systems. These oscillations are rationalized in terms of bifurcations of the eigenvalue problem  
and reflect the directed swimming motion of the active particles. 
In addition to the mean-square displacement, we have analyzed the non-Gaussian 
parameter and identified a characteristic maximum for positive anisotropies and 
large  P{\'e}clet numbers.

The non-Gaussian parameter has been derived before for two-dimensional 
isotropic swimmers~\cite{Sevilla:2014,Sevilla:2015} by a truncated 
mode expansion of the Fokker-Planck equation. Yet, for isotropic diffusion 
the non-Gaussian parameter remains negative for all times, in contrast to
experimental observations~\cite{Zheng:2013}. The mode expansion also yields
approximate expressions for the ISF which in 
principle also display oscillations in time for the two-dimensional case.     

In differential dynamic microscopy experiments for dilute suspensions of \emph{E.\! coli} bacteria in three dimensions
an oscillatory behavior for the ISF has 
been observed and analyzed approximately in terms of pure persistent swimming
motion~\cite{Poon:2016}. Our results predict that these oscillations fade out for large 
as well as small wavenumbers which should in principle be also measurable 
in the set-up. The motility parameters then can be extracted from the measured ISF relying on different
wavenumbers. 
The dynamics on small length scales is dominated by translational diffusion, 
at intermediate ones by the swimming motion, and, finally at large length scales 
by the rotational diffusion.   

Furthermore the spatio-temporal information obtained from the ISF allows to 
discriminate quantitatively the dynamics of different swimming behaviors,
whereas the mean-square displacement of several models such as simple run-and-tumble
motion~\cite{Martens:2012} is hardly distinguishable to that of an active Brownian particle. 

The analytic solution for the active Brownian swimmer derived here 
should serve as a reference for more complex swimming behavior. 
For example, \emph{E.\! coli} bacteria display a distribution of swimming
velocities, which can be accounted for directly by post-averaging our results
for the ISF. Similarly, the swimming velocity may fluctuate itself~\cite{Romanczuk:2012}
leading to a further smearing of the oscillations in the ISF. Furthermore, the rotational diffusion for 
bacteria should be complemented by a run-and-tumble motion~\cite{Berg:1972} as observed by particle 
tracking. Species-specific propulsion mechanisms, such 
as circular motion of the algae \emph{Chlamydomonas reinhardtii} \cite{Poon:2016}, can be accounted for
by introducing a torque in the Fokker-Planck equation. Our solution strategy can be adapted also
to two-dimensional systems, for instance for the movement of 
Janus particles~\cite{Zheng:2013} confined between two glass plates or for the 
circular motion of \emph{E.\! coli} bacteria close to surfaces~\cite{Lauga:2006}. 

\section*{Methods}
\subsection*{Expansion of the eigenfunctions in powers of the wavenumber}
The starting point of the expansion are the reference solutions 
$\text{Ps}_\ell^0(0,0,\eta)\equiv |\ell \rangle :=\text{P}_\ell(\eta)\sqrt{(2\ell+1)/2}$ for 
$\ell \in \mathbb{N}_0$ of the eigenvalue problem, Eq.~(\ref{eq:Pslm}), for parameters $R=c^2=0$. By standard perturbation theory one derives 
to the desired order $\mathcal{O}=\mathcal{O}(R^3,c^2R,c^4)$
\begin{align}
  \text{Ps}_\ell^0(R,c,\eta)	&= |\ell\rangle -R\biggl[|\ell-1\rangle \frac{\langle \ell-1|\eta|\ell\rangle}{\Delta A_\ell^{\ell-1}}
+|\ell+1\rangle\frac{\langle \ell+1|\eta|\ell\rangle}{\Delta A_\ell^{\ell+1}}\biggr]
+R^2\biggl[-\frac{1}{2}|\ell\rangle\left(\frac{|\langle\ell-1|\eta|\ell\rangle|^2}{(\Delta A_\ell^{\ell-1})^2}+
\frac{|\langle \ell+1|\eta|\ell\rangle|^2}{(\Delta A_\ell^{\ell+1})^2}\right)\label{eq:Psexpansion}\\
  & \ \ \ \ \ + |\ell-2\rangle \frac{\langle\ell-2|\eta|\ell-1\rangle\langle\ell-1|\eta|\ell\rangle}{\Delta A_\ell^{\ell-2}\Delta A_\ell^{\ell-1}}
 +|\ell+2\rangle \frac{\langle \ell+2|\eta|\ell+1\rangle\langle\ell+1|\eta|\ell\rangle}{\Delta A_\ell^{\ell+2}\Delta A_\ell^{\ell+1}}\biggr]\notag\\                       
  & \ \ \ \ \ +c^2\biggl[|\ell-2\rangle\frac{\langle \ell-2|\eta^2|\ell\rangle}{\Delta A_\ell^{\ell-2}}
+|\ell+2\rangle \frac{\langle\ell+2|\eta^2|\ell\rangle}{\Delta A_\ell^{\ell+2}}\biggr]+\mathcal{O},\notag 
\end{align} 
with corresponding eigenvalues 
\begin{align}
  A^0_{\ell}(R,c)	&= \ell(\ell+1)+R^2\left[\frac{|\langle\ell-1|\eta|\ell\rangle|^2}{\Delta A_\ell^{\ell-1}}
+\frac{|\langle \ell+1|\eta|\ell\rangle|^2}{\Delta A_\ell^{\ell+1}}\right] +c^2\langle \ell|\eta^2|\ell\rangle+\mathcal{O}.
\end{align}
Here $|\ell\rangle=0$ for $\ell<0$, 
the difference of unperturbed eigenvalues is denoted by $\Delta A_\ell^j = A^0_{\ell}(0,0)-A^0_{j}(0,0)$, 
and, the matrix elements of the perturbation 
\begin{align}
\langle n| \eta^j|\ell\rangle 	&= \sqrt{(2n+1)(2\ell+1)}\int_{-1}^1 \diff \eta \ \text{P}_n(\eta) \eta^j\text{P}_\ell(\eta)/2
\end{align}
for $j=1,2$ can be evaluated using the properties of the Legendre polynomials.

\subsection*{Numerical evaluation of the ISF}
For the ISF we need the eigenvalues $\text{A}^0_{\ell}$ and the integrals over the 
eigenfunctions $\text{Ps}_\ell^0$, Eq.~(\ref{eq:solISF}). We expand these in terms of the Legendre polynomials~\cite{Yan:2009}
  $\text{Ps}_{\ell}^0(c,R,\eta) = \sum\nolimits_{j=0}^\infty d^{0\ell}_j |j\rangle$. 
Then the integrals in Eq.~(\ref{eq:solISF}) can be performed and the intermediate scattering function of the anisotropic active Brownian particle reads 
\begin{align}
  F(k,t)  &= e^{-D_\perp k^2 t}\sum_{\ell=0}^\infty [d_0^{0\ell}]^2e^{-D_\text{rot}A^0_{\ell}t}, \label{eq:Fcoef}
\end{align}
Inserting the expansion 
into Eq.~(\ref{eq:Pslm}) 
and projecting onto $\langle n|$ leads to the matrix eigenvalue
problem 
\begin{align}
\sum_j[\langle n| c^2\eta^2-R\eta|j\rangle +n(n+1)\delta_{jn}]d_j^{0\ell} &= A^0_{\ell}d_n^{0\ell}.\label{eq:matrix}
\end{align} 

\begin{figure}[H]
  \centering
    \includegraphics[width=\linewidth, keepaspectratio]{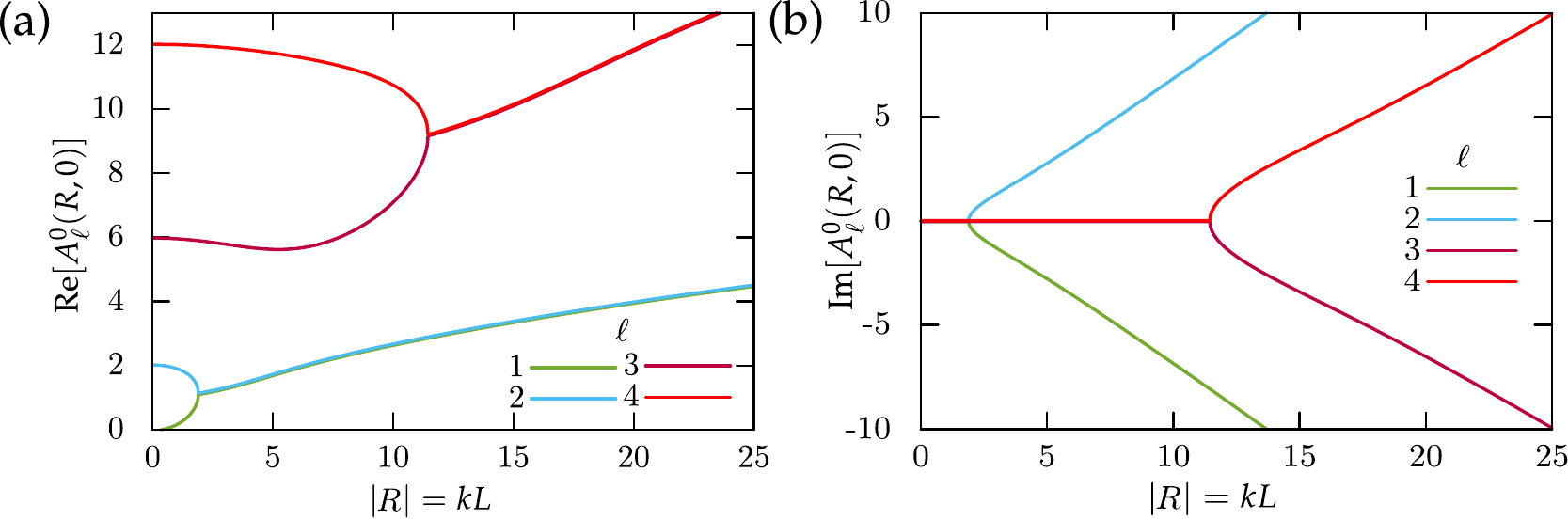}
     \caption{Real (a) and imaginary (b) part of the eigenvalues $A^0_{1}(R,0)$ to $A^0_{4}(R,0)$ for
     vanishing translational diffusion ($\text{Pe} =\infty$).\label{fig:bifEV}}
\end{figure}

Since the matrix elements are non-vanishing for $j=n-2,...,n+2$ only, 
it is in fact a band matrix with two diagonals on each side. Then the 
normalized eigenvectors $\vec{d}^{0\ell}= (d_0^{0\ell},d_1^{0\ell},d_2^{0\ell},...)^{\text{T}}$
and eigenvalues $A^0_{\ell}$ can be efficiently determined numerically. 
In practice we truncate the matrix in Eq.~(\ref{eq:matrix}) to sufficiently high order
such that the normalization at time $t=0$ for the ISF, Eq.~(\ref{eq:Fcoef}), is achieved. 
Since the generalized spheroidal wave equation is not
Hermitian, the corresponding eigenvalues can become complex.
In fact for $\text{Pe}=\infty$ ($c=0$), the two lowest eigenvalues 
merge at $|R|=kL=1.9$ and a bifurcation to two complex conjugates 
occurs, see Fig.~\ref{fig:bifEV}. In contrast for small P{\`e}clet number $\text{Pe}=1.1$
the eigenvalues remain real for all wavenumbers.




\begin{acknowledgements}
We acknowledge helpful discussions with Felix H\"ofling at the initial 
state of this project. This work has been supported by Deutsche 
Forschungsgemeinschaft (DFG) via the contract No. FR1418/5-1 and by the Austrian
Science Fund (FWF): P~28687-N27.
\end{acknowledgements}

\section*{Author contributions statement}

Author contributions: T.F. conceived the project.  S.L. designed
the simulation algorithm. C.K. implemented the theory and performed simulations. 
C.K. and T.F. interpreted the data and wrote the paper. All authors discussed the results and
commented on the manuscript.

\section*{Additional information}

\subsection*{Competing financial interests:} The authors declare no competing financial interests.

\end{document}